# Use of openMosix for parallel I/O balancing on storage in Linux cluster

**Gianluca Argentini**
*New Technologies and Models*
Information & Communication Technology Department,
Riello Group, Legnago (Verona) – Italy, via Alpini 1
*e-mail*: gianluca.argentini@riellogroup.com

**Abstract**

*In this paper I present some experiences made in the matter of I/O for Linux Clustering. In particular is illustrated the use of the package openMosix, a balancer of workload for processes running in a cluster of nodes. I describe some tests for balancing the load of I/O storage massive processes in a cluster with four components.*
*This work is been written for the proceedings of the workshop* "Linux cluster: the openMosix approach" *held at CINECA, Bologna – Italy, on 28 november 2002.*

**A brief presentation**

OpenMosix ([6]) is a free, open source, Linux kernel extension for single-system image clustering. It automatically balances the load between different nodes of the cluster.
CINECA ([7]) is the italian InterUniversitary Consortium for Automatic Calculus, at Casalecchio di Reno, Bologna. In this centre there are many systems for the parallel computing and in particular a Linux Cluster of 64 biprocessor nodes.
DEMOCRITOS ([8]) is the National Simulation Center of the Italian Istituto Nazionale per la Fisica della Materia, hosted by Scuola Internazionale Superiore di Studi Avanzati SISSA in Trieste. It has been, together with Cineca, the organizer of the workshop on openMosix.

**Introduction and definition of the discussed question**

The manifacturing concern Riello Group has centre at Legnago (Verona), produces domestic and industrial burners and markets conditioning equipments. The *Information & Communication Technology* department attends to computing procedures and infrastructures. The elaboration environment is distributed either geographically, being localized at three principle computation centres, or in the matter of hardware resources.
There are about 20 Unix servers of middle-high level and about 60 NT servers, and those used for the administration of the databases for ERP software and business Datawarehouse are multiprocessor with dedicated storage on each single server. The servers's total capacity of space on device is about 2 TeraBytes, distributed on hardware systems which are heterogeneous as technical features and available volume. This fact implies some troubles for the backup services, mainly implemented by three tape libraries, each with only one head for writing, and hence there are problems of performances and of serial queuing of the relative procedures running during the night. Furthermore in some servers a lot of programs, onerous in required time, in CPU waste and in I/O band on storage, necessaries for the administration of the databases and for the night population of the datawarehouse, run during the same temporal window. This fact and the increase of online data imply a progressive reduction of the temporal windows useful for the backup procedures. To find a remedy for this situation, even in a partial manner, some methods of defered startups by the Unix utility *cron* have been implemented for the working night-procedures with the aim of running one process for one processor, but the results have been modest.

For a better strategy, the ICT manager staff is planning for the triennium 2003-2005 a project of storage consolidation, with the aim of centralize the administration of device-space by a SAN for the Unix and NT servers; a project of server consolidation for the requirements of Erp and datawarehouse, and a pilot-project of a cluster Linux. In particular this latter resource will be used for the study and tests of parallel procedures either for business sphere (HA, high availability) as for technical and scientific computations (HPC, high performances computing) for fluidodynamical simulations. Furthermore, in consideration of some critical situations occurred the last summer 2002 as consequence of violent meteorological fenomena, the Linux cluster could be used for implementing local weather forecast with small temporal limits.

**The Linux cluster for tests**

For reducing the amplitude of the temporal windows devoted to night serial backups, we have planned a project to verify the possibility and the efficiency of rescue copies executed in parallel mode on device for db datafiles. We preferred this logic and not much expensive method to hardware implementations, less flexible and usually honerous. We remember that the aim was a study of fattibility, not the research of the best performances.

It has been built a small cluster with commodity components: 4 pc monoprocessor, three with Pentium II and one with Pentium III, respectively of 440 and 800 MHz, equiped in medium with a Ram of 256 MB. The computers have been connected to a private 100 Mb network, together with a NAS-like storage system, constituted by 4 disks for a total of about 120 GB; the storage system has been connected to the network by mean of two access channels.
The storage system has been configured for the pubblic sharing, by an NFS-like protocol, of two file systems of equal size.
On the nodes of the cluster it has been installed the operating system Linux, in the RedHat distribution release 7.2; the Oracle 8.1.6 engine and the two file systems exported by the NAS have been mounted at the point /dati and /copie. In every computer it has been built a small database with 2 datafiles of 2 GB, 2 of 1.5, 2 of 1 and 2 of 0.5, with the purpose of doing tests by mean of various-sized files. The tests have been organized for the simultaneous startup on the nodes of cluster for the copies of the files of same dimension, and on the contrary serializing on the same node the files copies for the disequal sizes. At the aim of simulating the load conditions, due to the running of the programs of db management and datawarehouse population, of our present business servers during the night, on the Pentium III node it has been scheduled the startup of some programs of elaborative weight, to be in execution simultaneously to the datafiles copy processes. This further requirement has induced us to consider the use of a software for balancing the total load of cluster, and we decided to test the *openMosix* package ([6]).

**First results for copy processes without openMosix**

As first tests, copy processes implemented by the system command *cp* have been used. On the Pentium III node simultaneously have been launched the simulation programs, which ran for about 1h 30m. The I/O registered with this method has been of about 20 minutes for each GigaByte on the

Pentium II nodes, hence for a total copy time of about 3h 20m for the 10 GB of the database of each single nodes, and of about 30m per GB on the Pentium III, for a total time of about 5h.

The result obtained with these first experiments has been a small degree of parallelism, for the reason that each process has copied only its data, and hence we have had not just a parallelism but a simultaneity of events, and a small balancing of the cluster load.

**Using openMosix**

Hence we have used the rel. 2.4.16 of openMosix to test the possibility of realize a better load balancing. To each copy process, simply implemented by the system command *cp* too, it has been associated a single datafile. The use of a single shared storage, external to all the nodes, has permitted to avoid the typical problems of cache consistency which rise up using NFS in the case of process migration from one node to another ([3]). The registered times in the presence of the copy programs alone have been of about 13 – 14 minutes per GB. Further some processes, at their startup, have been migrated by openMosix towards the Pentium III node, and this in particular after the copy of higher size datafiles in the Pentium II nodes, hence at a moment when the machine-load was considered even of big entity by the scheduling algorithm of openMosix system ([2]). The total load of the cluster has resulted well balanced.

Subsequently the experiment has been repeated with the simulation programs running on the major node. In this case it has been observed migration of copy processes from this node towards those less power, evidently considered by openMosix in the right condition for to balance the total load of cluster. Effectively the balance has been good, but the further load on the Pentium II nodes has degraded their performances, and the medium copy time for one GB is resulted of about 25 – 30 minutes. The Tab.1 offers a comparison of the first two experiments done on the use of openMosix.

|  | *No openMosix* | *With openMosix* |
|---|---|---|
| *Pentium II total time per node* | 3h 20m | 4h 30m |
| *Pentium III total time* | 5h | 4h |
| *Balancing* | small | good |

**Tab. 1**
*A comparison of the copy procedures with and without openMosix.*

From the table the following first conclusions can be derived:
- openMosix has reached the balancing of cluster nodes, that is the primary aim for which it has been projected;
- the functionality of the user-tools associated to the package, *mtop* for the load of a single node, *mps* for the control of a single active process, *mosmon* for the global monitoring of the cluster, has been good;
- the use of this package has implied in the described situation a saving, even if modest, of the time of total elaboration of copy processes;
- without the use of suitable file systems as MFS ([1], [2]), openMosix seems to privilege the balance of the load of computational activity on CPUs respect to that of I/O towards the storage system; this fact is proved by the migration of the copy processes from the Pentium III node, very CPUs-loaded by simulation programs, towards the other nodes, hence the saved hour of total elaboration on the powerful node has been lost as longer activity on Pentium II nodes.

**Using MPI**

For a better comprehension of the performances offered by openMosix, we have realized a copy procedure of datafiles by mean of the parallelization library *MPI*, Message Passing Interface, using the Argonne National Laboratories MPICH implementation, rel. 1.2.4 ([5]). For this aim it has been found as optimal technique for obtaining good performances the use of multiprocess views, that is the possibility offered by some specific functions of the package MPI-2 of reading and writing a fixed portion of a single file for each process (see Fig.1, [4]). The copy procedure has been wrote in C language and compiled by mean of GNU C rel. 2.3 .

A remarkable characteristic of this library is the possibility of using a buffer for reading or writing, and after some tests it has been chosen one of 25 MB size for each node. A bigger value could be better for a faster I/O but it could cause overflow of nodes memory.

Some tests have been conducted with 1, 2 and 4 processes for each datafiles, registering the copy times even for every single files size. It has not made use of 3 processes for avoiding problems of no perfect divisibility of the file size by multiples other then those of 2.

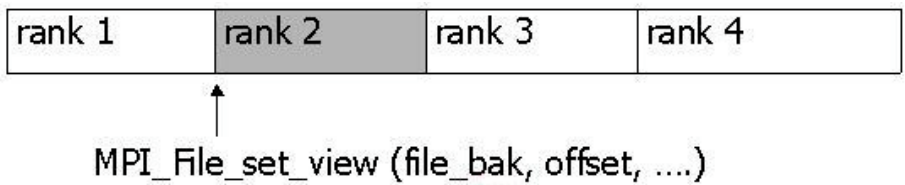

**Fig. 1** *A datafile segment with views for each process, identified by its own rank (ID code in MPI). The process number 4 is shown with a view of size not equal to the other because of the possibility of a non integer division by 4 between the datafile size, the buffer size, and the number of segments necessary for the covering of the entire file.*

Using only the MPI library and not openMosix, the medium time for the copy of 1 GB on Pentium II nodes has been of 13 – 15 minutes, and of 15 –17 on Pentium III where ran simultaneously the simulation programs too. The complete load of the cluster has been satisfactory even if not optimized. Using instead openMosix too, some migrations of copy processes from Pentium III to the other nodes has been risen up; this fact seems to confirm the CPU-oriented nature of the internal algorithms of openMosix, which tends to balance with priority the computational load on processors. During the temporal period of processes migration, the cluster general load has been very well balanced, but the Pentium II nodes, which were heavier respect to the previous case, have registered an increment of about 25% of their local copy time. Instead the total backup time on the cluster is remained almost the same. The following Tab.2 offers a summary of the executed experiments:

|  | *Only openMosix* | *Only MPI* | *MPI-openMosix* |
|---|---|---|---|
| *Pentium II total time* | 4h 30m | 2h 30m | 3h |
| *Pentium III total time* | 4h | 2h 50m | 2h 30m |
| *Balancing* | good | sufficient | very good |

**Tab. 2** *Comparison of the copy procedures and simulation programs times between openMosix and MPI.*

**Comparison tests between openMosix and MPI**

To obtain some results on the degree of parallelism and on the balancing of the load by mean of a comparison between openMosix and MPI, has been executed three copy experiments with a node datafiles set of about 8 GB, and respectively with size of 0.4 GB (20 files), 1 (8), 1.4 (6), 2 (4). On every node have been run, with openMosix started (*openmosix start*), at first *cp* copy processes simultaneously in number of 1, 2 and 4 on three successive experiments, and then on the cluster, with openMosix stopped (*openmosix stop*), have been launched copy procedures managed by MPI with 1, 2 and 4 processes per node on other three experiments.

We have not used on the Pentium III the simulation programs for not to dirty the test making the results difficult to analyze. In this manner on the cluster at each instant was running, in the two situations, the same number of processes, which have acted on every node on the same data quantity. Hence the two analized situations have been based on two distinct philosophies: with openMosix we attempted to parallelize the total procedure on the cluster associating to each process one file for the copy, with MPI associating to each file one multiprocess program. The reached conclusions, graphically shown in Fig.2, can be so summarized:

- openMosix has registered a less copy time in the case of one process per node and in the multiprocess case for small size files;
- openMosix has shown a performances degrade increasing respect the used processes number, demonstrating that the load of the autonomous concurrent processes on the resources of the single nodes has had a negative weight bigger than the obtained parallelism;
- for MPI the use of its native parallelism and the possibility of buffers management have shown an increase of performances, even if with decreasing influence, respect the number of used processes.

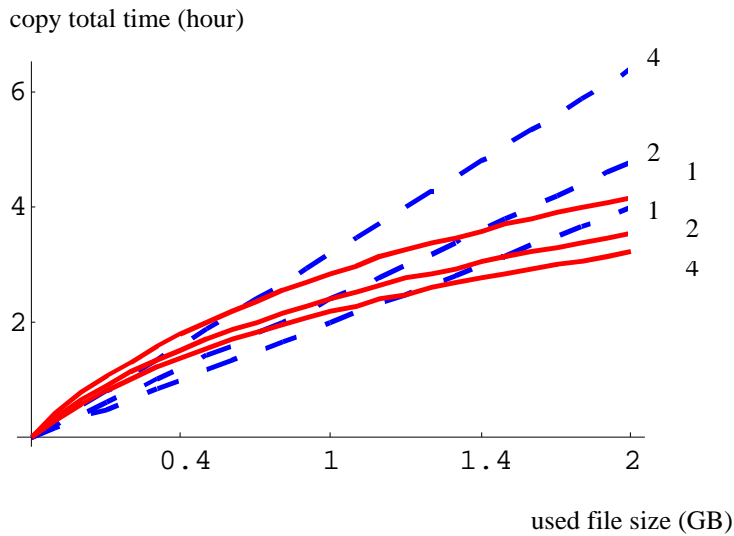

**Fig. 2**
*Comparison between the copy times of openMosix (dashed lines) and MPI. The numbers beside the lines correspond to the processes used for obtain the graphic. The growth of MPI temporal graphics is resulted to be logarithmic respect the used files size.*

**Final considerations**

The used cluster Linux has had evident limits: diversity of the nodes hardware characteristics; the nodes itself were only monoprocessor; the original files to be read and those to be written were localized on the same disks set; it has not been possible to conduct tests with files of size bigger than 2 GB, for an intrinsic limit of the used storage system; it has not been possible to use the file system MFS, Mosix File System, which in analogous studies has provided very good results with openMosix ([1]). From the executed tests the following considerations can be extracted:
- openMosix has evidenced correct and good functionalities of processes balancing;
- its use with MPI has not reduced the cluster global performances of the library;
- it has shown inferior results compared to MPI in the case of multiprocessor procedures parallelizing the I/O of big size files.

**Acknowledgments**

The author wish to thank *Stefano Martinelli* and *Carlo Cavazzoni* of CINECA for the invitation to present this work and for useful explanations on the use of functionalities of MPI-2; *Giorgio Colonna*, Riello Group ICT Department manager, and *Guido Berardo*, ICT Projects Planning manager, for to allow me the studies on HPC and Linux Clustering; *Ernesto Montagnoli*,

**References**

[1] Roberto Aringhieri, *Open source solutions for optimization on Linux clusters*, University of Modena and Reggio Emilia, DISMI, Technical Report 23/2002;
[2] Moshe Bar, *openMosix Internals*, article in openMosix Web site;
[3] Moshe Bar, *Oracle RAC on Linux*, BYTE, July 2002;
[4] William Gropp – Ewing Lusk, *Advanced topics in MPI programming*, in T. Sterling, Beowulf cluster computing with Linux, 2002, MIT Press;
[5] www.mcs.anl.gov/mpich, Web site for MPI libraries;
[6] www.openmosix.sourceforge.net, Web site for openMosix;
[7] www.cineca.it, Web site for Cineca;
[8] www.democritos.it, Web site for Democritos.